\RequirePackage[hyphens]{url}
\documentclass[pageno]{jpaper_custom}

\usepackage{algorithm}
\usepackage{algpseudocode}
\usepackage{amsmath}
\usepackage{amssymb}
\usepackage{balance}
\usepackage{bbding}
\usepackage{booktabs}
\usepackage{bookmark}
\usepackage{comment}
\usepackage{datetime}
\usepackage{enumerate}
\usepackage{enumitem}
\usepackage{framed}
\usepackage{graphicx}
\usepackage{inputenc}
\usepackage[procnames]{listings}
\usepackage{listings}
\usepackage{microtype}
\usepackage{multirow}
\usepackage[all]{nowidow} 
\usepackage{relsize}
\usepackage{setspace}
\usepackage[nounderscore]{syntax}
\usepackage{textgreek}
\usepackage[normalem]{ulem}
\usepackage{url}
\usepackage{xcolor}
\usepackage{xspace}
\usepackage{afterpage}

\newcommand{\SCCLang}{GC3\xspace}
\newcommand{\msft}{Microsoft\xspace}
\newcommand{\scclef}{GC3-IR\xspace}

\newcommand{\rankdag}{Instruction DAG\xspace}
\newcommand{\chunkdag}{Chunk DAG\xspace}

\newcommand{\alltonextspeedup}{$14.5\times$\xspace} 
\newcommand{\alltoallscclvhand}{$1.3\times$\xspace} 
\newcommand{\alltoallscclvnccl}{$20\%$\xspace} 
\newcommand{\allreducering}{$1.9\times$\xspace}

\newcommand{\allreducehierarchical}{$1.4\times$\xspace}
\newcommand{\allreducehierarchicallarge}{$11\%$\xspace} 

\lstdefinestyle{mystyle}{
    language=Python,
    basicstyle=\footnotesize\ttfamily,
    breakatwhitespace=false,         
    breaklines=true,                 
    captionpos=b,                    
    keepspaces=true,                 
    numbers=left,                    
    numbersep=1mm,    
    numberstyle={\tiny},              
    showspaces=false,                
    showstringspaces=false,
    showtabs=false,                  
    tabsize=1,
    columns=flexible,
    keepspaces,
    xleftmargin=1em,
    basicstyle=\footnotesize\ttfamily,
    keywordstyle=\color{blue!60},
    keywordstyle=[2]\color{orange!60!black},
    keywordstyle=[3]\color{green!60!black},
    stringstyle=\color{purple},
    commentstyle=\color{green!70!black},
    procnamestyle=\color{blue},
    lineskip=-1em,
    escapechar=`,
    upquote=true,
    morekeywords=[2]{reduce,copy,chunk},
    morekeywords=[3]{parallelize,ch,sendtb,recvtb},
    deletekeywords=[2]{next,buffer},
}
\usepackage{textcomp}
\usepackage{tabularx}
\begin{document}

\title{\SCCLang: An Optimizing Compiler for GPU Collective Communication
}

\date{}

\author{
{\rm Meghan Cowan}\\
Microsoft Research
\and
{\rm Saeed Maleki}\\
Microsoft Research
\and
{\rm Madanlal Musuvathi}\\
Microsoft Research
\and
{\rm Olli Saarikivi}\\
Microsoft Research
\and
{\rm Yifan Xiong}\\
Microsoft Research Asia
} 

\maketitle

\begin{abstract}
Machine learning models made up of millions or billions of parameters are trained and served on large multi-GPU systems.
As models grow in size and execute on more GPUs, the collective communications used in these applications become a bottleneck.
Custom collective algorithms optimized for both particular network topologies and application specific communication patterns can alleviate this bottleneck and help these applications scale.
However, correctly and efficiently implementing custom algorithms is challenging. 

This paper introduces \SCCLang, a system for programmable GPU communication.
\SCCLang provides a domain specific language for writing collective communication algorithms and an optimizing compiler for lowering them to an executable form, which can be executed efficiently and flexibly in an interpreter based runtime.
We used \SCCLang to write novel collective algorithms for AllReduce and AllToAll that are up to \allreducering and \alltoallscclvhand faster than hand-optimized implementations, respectively.

\end{abstract}

\section{Introduction}
\label{sec:introduction}
Recent trends in machine learning (ML) point towards model sizes growing at a much faster rate than a single GPU's memory capacity and computational power~\cite{exp1, exp2}. This necessitates distributing model parameters across multiple GPUs~\cite{megatron,nvidiagpt3,deepspeedmultigpuinference} and associated communication to synchronize activations and/or model weights. As models become larger, the cost of communication increases as a percentage of total GPU execution time. For instance, training Resnet50~\cite{resnet} with $\approx$100MB of parameters spends 3\% of the time in communication~\cite{nsdisapio}, while training DeepLight~\cite{deeplight} with $\approx$2GB of parameters spends 79\% of its time in communication on the same distributed system.
Moreover, as models become larger, distribution is necessary not only for training, but also for inference. 

GPU communication kernels implement MPI {\em collective} communication operations, such as AllReduce, AllGather, and AllToAll~\cite{orgmpi}. Implementing these collectives requires communication algorithms, such as the Ring and Tree algorithms~\cite{thakur2005optimization}. Recent research~\cite{cho2019blueconnect,wang2020blink,cai2021synthesizing,xie2021synthesizing} has shown that custom algorithms that are specifically optimized for underlying interconnection topology and input sizes result in significant performance improvements. 

Communication kernel implementations make trade-offs between performance and algorithmic flexibility. On one hand, vendor libraries such as NCCL~\cite{ncclrepo} and RCCL~\cite{rcclrepo}, provide high-performance implementations of a few standard algorithms, such as Ring and Tree. These manually-written kernels implement a variety of low-level optimizations, such as tiling, parallelization, and fusion that are carefully tuned for performance. However, manually writing and optimizing these kernels is error prone due to many correctness pitfalls, such as deadlocks and data races. Repeating this process for the multitude of custom algorithms is extremely resource intensive. 

On the other hand, prior works that explore the space of custom algorithms do not provide efficient implementations for these algorithms. Partly to avoid the complexity of implementing custom kernels, many of the works~\cite{cho2019blueconnect,wang2020blink,xie2021synthesizing} compose existing CUDA kernels. These approaches not only incur the cost of multiple kernel-launches, they lose the opportunity to perform optimizations that cross kernel boundaries. Cai et~al.~\cite{cai2021synthesizing} generate customer kernel using low-level send, receive, and reduction operations~\cite{cai2021synthesizing}, but do not implement the low-level optimizations that high-performance vendor libraries do.  

This paper proposes \SCCLang, a unified framework that provides both algorithmic flexibility and performance.  
\SCCLang is made up of a {\em domain-specific language (DSL)} for specifying communication algorithms, a {\em compiler} for generating high-performance executables from these high-level specifications, and an efficient {\em runtime} for \SCCLang executables. 
The key contribution of this paper is that for a given collective communication algorithm, a developer can explore different implementations and optimizations in \SCCLang's DSL without fearing data races/deadlocks or writing any C/CUDA code while enjoying the performance of a hand-written code. Additionally, \SCCLang automatically checks whether the implementations has the intended semantics of the collective. Lastly, the runtime is API-compatible with NCCL allowing existing ML workloads to easily switch over to \SCCLang, inherit NCCL's support of diverse set GPUs and inter-connections, and safely fall over to NCCL kernels for yet unimplemented algorithms in \SCCLang to enable safe operation in production. 
\SCCLang is publicly available: \url{https://github.com/microsoft/msccl-tools} and \url{https://github.com/microsoft/msccl}.

\begin{figure*}[h!]\centering
    \begin{minipage}[b]{0.74\textwidth}
        \includegraphics[width=0.9\textwidth]{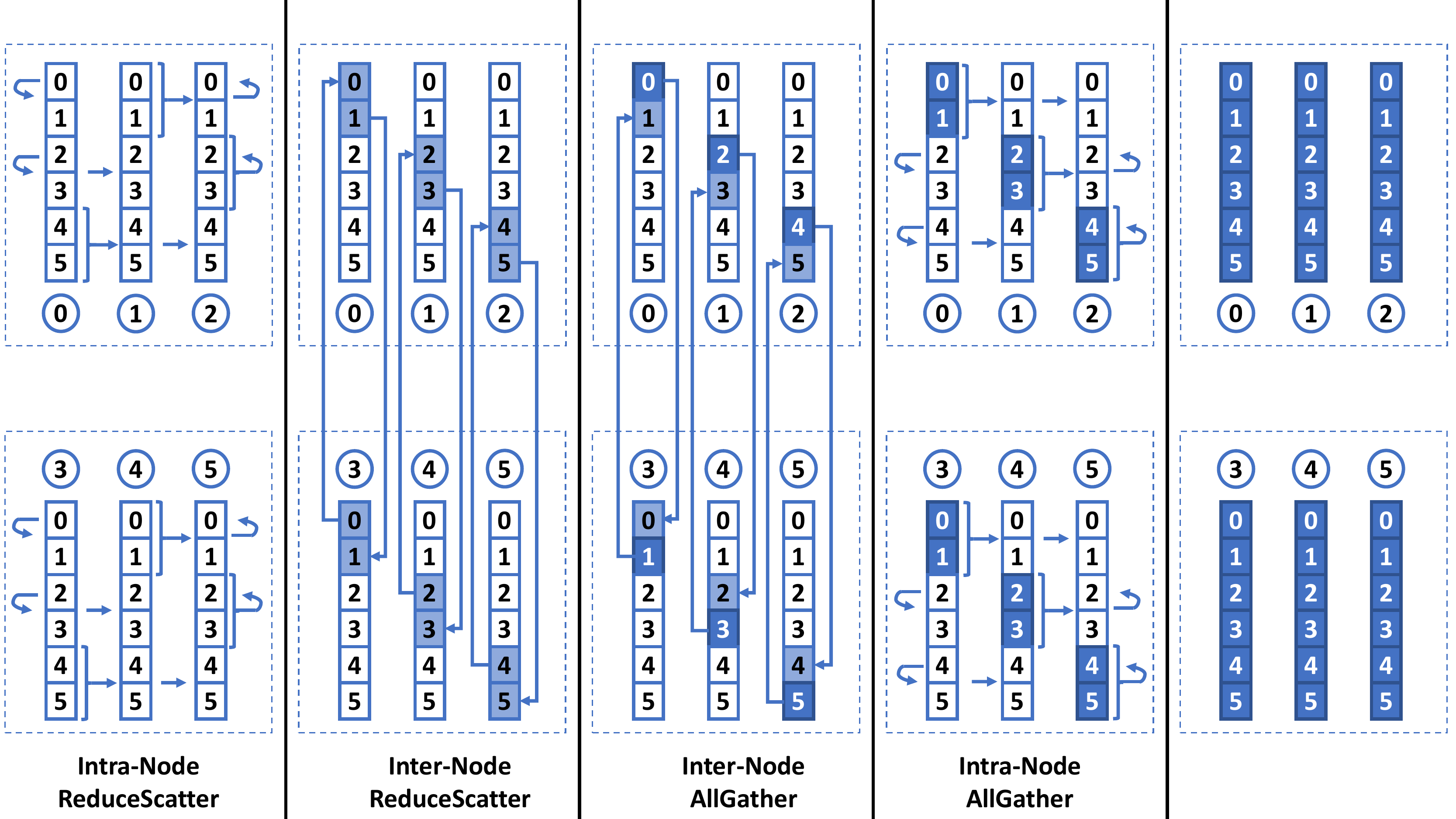}
        \caption{Hierarchical AllReduce on 2 nodes each with 3 local GPUs.}
        \label{fig:hier_allreduce_animation}
    \end{minipage}%
    \begin{minipage}[b]{0.26\textwidth}
        \centering
        \includegraphics[width=0.9\textwidth]{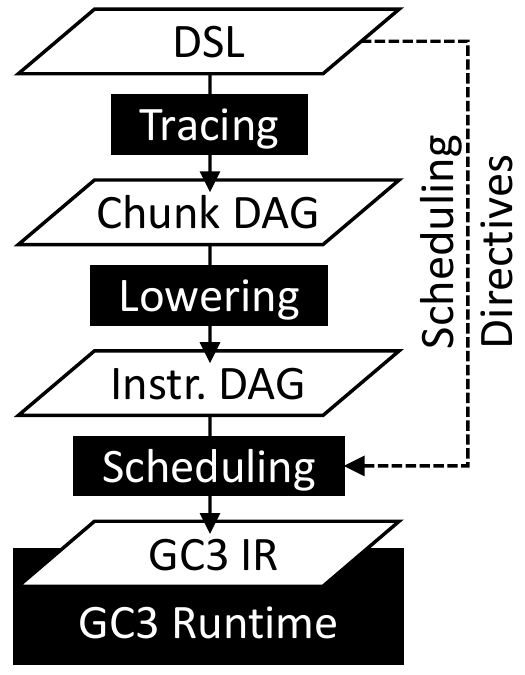}
        \vspace{2em}
        \caption{Architecture of \SCCLang{}\label{fig:architecture}}
    \end{minipage}
\end{figure*}

We evaluate \SCCLang  on two distributed GPU systems: a cluster of 8-A100 nodes and a cluster of 16-V100 nodes. We show that for a given algorithm, \SCCLang implementations match, and often beat, the performance of a hand-written implementation. 
This includes an AllToAll algorithm on multiple nodes that is up to \alltoallscclvhand faster than a hand-optimized implementation and Ring AllReduce algorithm that is up to \allreducering faster than NCCL's optimized implementation.
Additionally, we make a case for custom collectives by replacing simple point-to-point communication with a new collective called AllToNext. 
Lastly, \SCCLang system is used in inferencing a public facing language model on 8xA100 GPUs and training a large Mixture-of-Experts model for speech, language, and vision on 256xA100 GPUs at \msft providing 1.22-1.29$\times$ and 1.10-1.89$\times$ speed up, respectively.
\section{\SCCLang Example}
\label{sec:overview}

This section introduces \SCCLang through a running example: Hierarchical AllReduce. 
\autoref{fig:hier_allreduce_animation} shows the working of this algorithm. For a topology of $N (=2)$ nodes and $G (=3)$ GPUs per node, the algorithm splits the input into $N \times G (=6)$ chunks. The algorithm proceeds in four phases. The first phase is an intra-node ReduceScatter that computes the sum of buffers within a node but the result is "scattered" across the GPUs. This is done through a Ring algorithm in the figure. GPU $1$ sends $N$ chunks (chunk $0$ and chunk $1$) to GPU $2$ which adds them to its corresponding chunks before sending them to GPU $0$. In the end, GPU $0$ has the intra-node sum of these $N$ chunks, which is shown as lightly shaded in the figure. Other GPUs have intra-node sum of $N$ other chunks each by executing a similar ring as shown in the figure. The second phase is an inter-node ReduceScatter, where GPUs with the same intra-node sum chunks communicate across nodes. For instance, GPU $0$ and GPU $3$ use a Ring algorithm to add the intra-node sums of chunk $0$ and chunk $1$. The result is scattered with each GPU having one chunk of the AllReduce result, which is shown as darkly shaded in the figure. The final two phases are an inter-node AllGather followed by an intra-node AllGather, both of which follow a similar Ring algorithm to distribute these chunks to all GPUs.

\begin{figure*}[t]
    \centering
    \begin{subfigure}[b]{0.5\textwidth}
        \input{text/hierallreduce_program_nochan.tex}
        \caption{\SCCLang program for hierarchical AllReduce.}
        \label{fig:hier_allreduce_skeleton}
    \end{subfigure}%
    \begin{subfigure}[b]{0.5\textwidth}
        \input{text/redscatallgat_program_nochan.tex}
        \caption{\SCCLang helper functions for Ring ReduceScatter and AllGather.}
        \label{fig:red_scatter_all_gather}
    \end{subfigure}
    \caption{\SCCLang programs for hierarchical AllReduce.}
    \label{fig:hier_allreduce}
\end{figure*}
 
\paragraph{\SCCLang Program} 
\SCCLang has a DSL embedded in Python to write communication algorithms by declaratively specifying chunks  routes across the GPUs to implement a collective. We call such specifications {\em chunk-oriented}. \autoref{fig:hier_allreduce} shows an example for the hierarchical AllReduce algorithm. When interpreted as a Python program, the execution mimics the description in \autoref{fig:hier_allreduce_animation}. \autoref{fig:hier_allreduce_skeleton} creates the four phases: $N$ and $G$ instances of intra-node and inter-node ReduceScatter and AllGather, respectively. \autoref{fig:red_scatter_all_gather} implements ReduceScatter and AllGather using the Ring algorithm. In \SCCLang, a {\tt chunk} is identified by its rank and its index into a buffer in the rank, as shown in \autoref{line:redscat_chunk} and \autoref{line:allgat_chunk}.  As this chunk is routed across the ring, ReduceScatter performs {\tt reduce} at \autoref{line:redscat_op} while AllGather performs a {\tt copy} at \autoref{line:allgat_op}. \autoref{sec:language} explains the \SCCLang DSL in detail.

\paragraph{\SCCLang Architecture}
\autoref{fig:architecture} describes the components of the \SCCLang framework. Given a \SCCLang program, the \SCCLang compiler traces the program to capture the chunk dependencies in a Chunk DAG. The compiler then performs several optimizations such as aggregation, tiling, parallelization and schedules the resulting chunk operations to thread blocks specified in the \scclef. \SCCLang DSL also allows users to control the optimizations and scheduling choices performed by the compiler. As an example, the hierarchical AllReduce switches between intra-node communication and inter-node communication. \SCCLang uses tiling to automatically generate multiple instances of AllReduce each operating on a smaller chunk and pipelines these instances to improve link utilization.  The \SCCLang runtime executes \scclef{} as a single CUDA kernel, while performing additional optimizations. 
The compiler ensure that distributed execution correctly implements the chunk-oriented semantics of the input program while guarantees the absence of deadlocks and data races. 

The key advantage of \SCCLang is that users get algorithmic flexibility to specify custom communication algorithms in a high-level DSL while still getting the performance of hand-written kernels. This paper compares the performance of several algorithm in \SCCLang against high-performance NCCL baseline.  
\section{\SCCLang DSL}
\label{sec:language}

The \SCCLang DSL is a chunk-oriented language for specifying chunk routing through GPUs. The language is embedded in Python as a traced DSL with a fluent API that
gives users flexibility in how to express algorithms. In this section, we explain the core components of the
\SCCLang DSL for chunk routing. In \autoref{sec:schedules} we discuss scheduling extensions to further 
optimize programs.

\subsection{Buffers and Chunks}
GPU memory is exposed to \SCCLang as named \emph{buffers},
three of which are available on each rank: \emph{input},
which starts out holding the input data, \emph{output} and \emph{scratch}, which are uninitialized. The job of the
\SCCLang program is to ensure that the output buffer on each GPU gets filled with the correct result for the collective.
The scratch buffer can be used by \SCCLang programs for temporary storage. 

Buffers are divided into \emph{chunks}, which represent contiguous spans of elements with a uniform size across the whole \SCCLang program. The number of chunks is explicitly specified in the \SCCLang program.
However, at the program level the buffer size is abstract; the size is known at runtime when the concrete buffers are passed into the program.
Chunks can take three forms:
\begin{itemize}
    \item \emph{Input chunks} represent chunks initialized at runtime and are uniquely identified by the (rank, index) of the input buffer they start at.
    \item \emph{Reduction chunks} are created by combining two chunks through a point-wise reduction (e.g. addition). Reduction chunks are identified by a list of input chunks that have been combined.
    \item \emph{Uninitialized chunks} are a unit type that stores uninitialized data. At the start of the program, the output and scratch buffers hold uninitialized chunks.
\end{itemize}

\subsection{Collective}
\SCCLang programs are associated with a \emph{collective} that defines a \emph{signature} and a
\emph{postcondition} any algorithm must guarantee. The signature determines the number of chunks in the input and output buffers for each rank. 
The signature also indicates if the program is \emph{in-place} and expects the input and output buffer alias each other.
For example, \autoref{fig:hier_allreduce} implements an in-place AllReduce with $N \times G$ chunks for input and output buffers.

The number of chunks in the scratch
buffers are not determined by the collective, but instead \SCCLang deduces it from the program based on the highest
indices accessed, which allows users to conveniently access extra memory.

The postcondition of a collective allows \SCCLang to automatically validate that a prospective algorithm implements its collective. For each index of each output buffer, the postcondition specifies either an input chunk or a reduction chunk that the algorithm must place into it. 

\subsection{\SCCLang Operations}
\label{sec:programs}

\begin{table*}[t]
    \centering
    \begin{tabularx}{\linewidth}{@{}lX@{}}
    \toprule
    \textbf{Operation} & \textbf{Description} \\ 
    \midrule
    \texttt{chunk(rank, buffer, index, count=1) $\to$ c} & 
        Returns a reference (rank, buffer, index, count) for the chunks currently in the buffer. \\
        \midrule
    \texttt{c1.copy(rank, buffer, index) $\to$ c2} &
        Copies chunks referenced by c1 into the destination indices.
        Returns a new reference c2 for the copied chunks. \\
        \midrule
    \texttt{c1.reduce(c2) $\to$ c3} & 
        Reduces chunks referenced by c1 and c2 \emph{in-place} into the indices of c1.
        Returns a new reference c3 for the result. \\
    \bottomrule
    \end{tabularx}
    \caption{\SCCLang DSL operations.}
    \label{tbl:scclang}
\end{table*}

\autoref{tbl:scclang} lists the \SCCLang operations used for manipulating chunks, which we discuss below.
Buffer indices are accessed with the free function \texttt{chunk(rank, buffer, index, count=C)} which returns a reference to \texttt{C} contiguous chunks currently assigned to the named \texttt{buffer} starting at \texttt{index}.
\texttt{count} default value is one when it is not set.
For correctness, \SCCLang raises an error if any of the chunks accessed are uninitialized.

Chunks are moved between buffers with the \texttt{copy} and \texttt{reduce} operations. Specifically,
\texttt{c1.copy(rank2, buffer, index2)} copies the chunks referenced by c1 to (rank2, buffer, index2).
Equal count chunk references, c1 and c2, can be reduced into a single result with \texttt{c1.reduce(c2)}, which
combines the chunks with an in-place point-wise operation that overwrites c1 with reduced chunks.
Copy and reduce return references to the newly created chunks, which allows fluently chaining \texttt{copy} and \texttt{reduce} calls.

Programs manipulate references rather than chunks to prevent operations on stale data.
Throughout the program, multiple references to the same (rank, buffer, index) location can be created and potentially refer to stale chunks that are overwritten by later operations.
\SCCLang only allows the latest reference for any location to be used and will generate an error otherwise.
This enforces a chunk-oriented coding style where the program script must always operate on current references, which makes \SCCLang programs data race free by construction. 

\SCCLang lets users express operations between buffers uniformly with \texttt{copy} and \texttt{reduce} regardless of
whether they are on the same GPU or not. The next section explores how \SCCLang enables this abstraction.

\section{Lowering \SCCLang Programs}
\label{sec:compiler}

\begin{figure*}[t]
    \centering
    \includegraphics[width=\linewidth,height=\textheight,keepaspectratio]{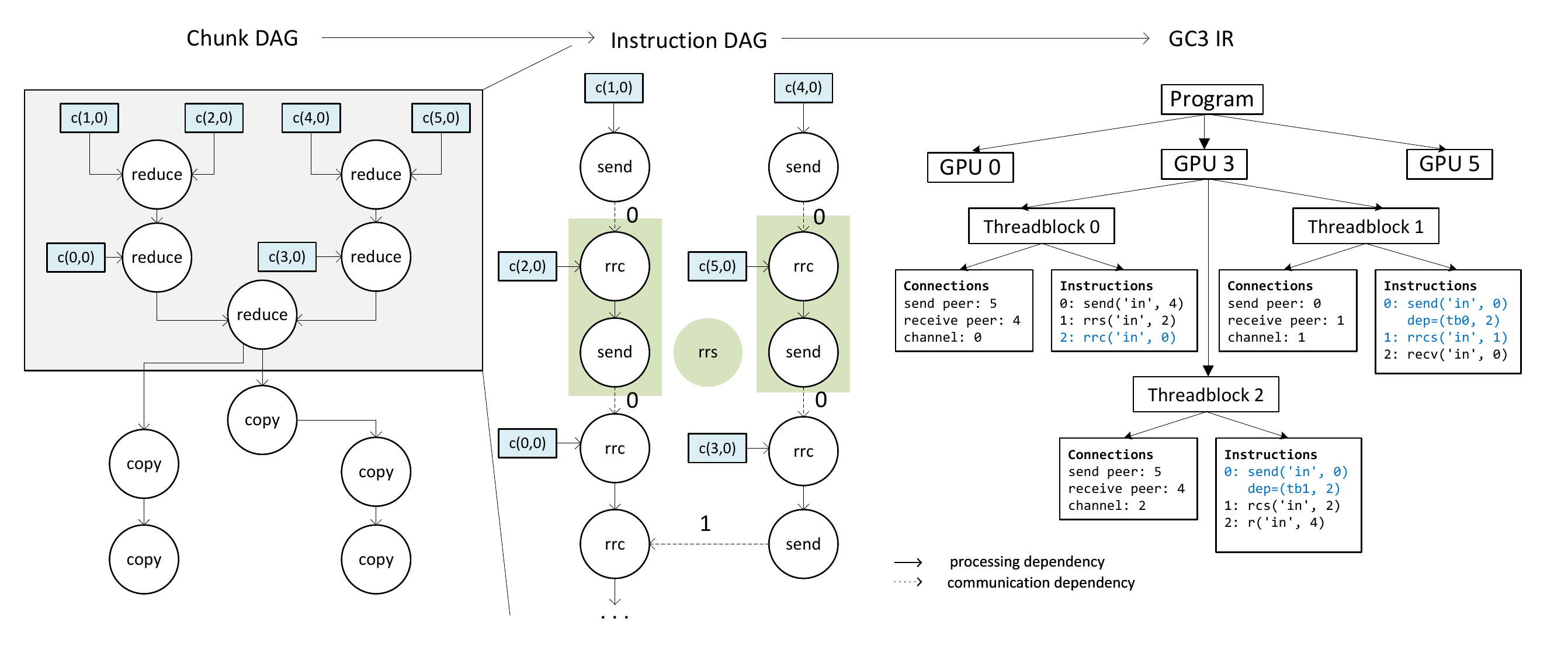}
    \caption{\SCCLang compilation of the hierarchical AllReduce.
    A subset of the program for chunk 0 is traced into a \chunkdag of operations, lowered 
    into an \rankdag of instructions, and then scheduled onto thread blocks.} 
    \label{fig:compiler}
\end{figure*}

The functions in \autoref{tbl:scclang} implement a tracing frontend that records a directed acyclic graph (DAG) of operations, which we call the \chunkdag. This is further lowered into an \rankdag, and finally scheduled into procedural code targeting the low-level \scclef for our runtime. 
This section explains how programs are lowered, \autoref{sec:schedules} discusses how programs are scheduled, and the optimization interface the DSL provides. 

\subsection{Tracing}

A \SCCLang{} program is sequentially traced into a \chunkdag, which captures the global view of chunk movement  and naturally exposes the program's parallelism. The graph is initialized with source nodes for all input chunks. Every \texttt{copy} and \texttt{reduce} operations is represented as a node, and the edges between nodes are dependencies that arise from chunk movement (true dependencies) and reusing buffer indices (false dependencies). 

\autoref{fig:compiler} depicts a subset of the  \chunkdag of \autoref{fig:hier_allreduce} that traces chunk 0 across every rank. The \chunkdag preserves the hierarchical structure of the program, with two levels of reduces required for the ReduceScatters and two levels of copies for the AllGathers.

\subsection{Instruction Generation}
Each chunk operation node is expanded into instruction nodes to generate the \rankdag. Instructions are either point-to-point communication primitives or local primitive that are executed by a single GPU. The instructions are listed below:

\begin{itemize}
    \item \texttt{send(buffer,index)/recv(buffer,index)}: 
    sends/receives from the given buffer at 
    the chunk index to/from the remote GPU.
    \item \texttt{reduce(srcBuf,srcInd,dstBuf,dstInd)}: 
    locally applies a pre-defined reduction operation to the corresponding chunks and stores the result in the destination.
    \item \texttt{copy(srcBuf,srcInd,dstBuf,dstInd)}: locally
    copies a chunk from a source location to a destination.
    \item \texttt{recvReduceCopy/rrc(srcBuf,srcInd,dstBuf,\\
    dstInd)}: 
    is a fused instruction that receives a chunk, reduces it 
    with a source chunk, and locally copies it to the destination.
    \item \texttt{recvReduceCopySend/rrcs, recvReduceSend/rrs, 
    recvCopySend/rcs}: are additional fused instructions. 
\end{itemize}
Fused instructions can be implemented by composing \texttt{send}, \texttt{recv}, \texttt{reduce}, and \texttt{copy} instructions. However, fusing them optimizes the global memory accesses as intermediate values are transferred through GPU registers. 

A chunk operation expands differently depending on whether it is local or remote. A remote \texttt{copy} expands into a \texttt{send} and a \texttt{receive} instruction, and a remote \texttt{reduce} expands into a \texttt{send}
and a \texttt{receiveReduceCopy} instruction. For local \texttt{copy} or \texttt{reduction} operations only a
single local instruction is generated. Note that instructions such as \texttt{receiveCopySend} cannot be generated this way as it requires looking at two chunk operations.

The two instructions resulting from a remote operation are connected by a communication edge, which represents the fact
that the receiving side synchronizes with the sender. The original edges of the \chunkdag are preserved as processing
edges, which represent the execution-order dependencies within ranks.

\paragraph{Instruction Fusion.}
Naive instruction generation can only use a subset of the available instructions excluding the fused instructions that combine a receive and a send. The compiler performs a series of peephole optimizations to combine consecutive base instructions into fused instructions. 

\paragraph{\texttt{rcs}}
Rewrites a back-to-back receive and send of the same size on the same buffer index into a fused \texttt{receiveCopySend}:

If there are multiple sends dependent on the receive, the send on the longest path in the \rankdag is fused.

\paragraph{\texttt{rrcs}} A back-to-back pair of a \texttt{receiveReduceCopy} and a \texttt{send} instruction are
rewritten into a \texttt{receiveReduceCopySend}. 

\paragraph{\texttt{rrs}} Is a special case of the previous optimization, if the result of the \texttt{rrc} is never used locally (i.e. it is later overwritten), the reduction result does not need to be saved locally and a more efficient \texttt{receiveReduceSend} instruction can be used instead.

~\autoref{fig:compiler} depicts the hierarchical AllReduce \rankdag for chunk 0 up to the inter-node ReduceScatter. Each operation node has been expanded into two instruction nodes, with communication edges connecting a matching send and receive. 
Highlighted in green, is a back-to-back \texttt{rrc} and \texttt{send} that is fused into a \texttt{rrs} instruction. 

\section{Scheduling \SCCLang Programs}
\label{sec:schedules}
After a program is lowered into an \rankdag, it is scheduled into a \scclef program that specifies how the program is executed. At a high-level this process assigns every instruction to a thread block that will execute it and every communication edge to a channel identifying the connection data is transferred through.

\paragraph{\scclef}
\scclef, captures scheduling decisions in a tree data structure (see ~\autoref{fig:compiler}).
A \scclef is divided into GPU programs that are subdivided into thread blocks containing a list of instructions that are sequentially executed.

Thread blocks can make two uni-directional connections to a \textit{send peer} and a \textit{receive peer} that are used for sending and receiving chunks.
GPUs are allowed to have multiple redundant connections to the same GPU that are identified by a \textit{channel}.
Channels are similar to tags in MPI that identify the route a send takes.

Our design restricts thread blocks to have at most one send and receive connections so that two thread blocks do not serialize over the same connection. Similarly, a connection can only have one sending and receiving thread block. The compiler ensures this constraint is honored by during scheduling.

\subsection{DSL Scheduling Directives}
Optimizing a program's schedule is crucial for extracting performance.
The \SCCLang DSL provides a set of \textit{scheduling directives} that can be used to specify optimizations on top of a program.
These optimizations trade off between parallelization, by assigning instructions across multiple thread blocks and occupancy constraints, as each thread block consumes GPU resources.

\paragraph{Channel Directives}
A program may use multiple connections between the same pair of GPUs that are differentiated by their channels.
Programs can specify that an operation utilizes a particular channel with an optional parameter \texttt{ch}.
The code below schedules two copies on different channels: 
\begin{lstlisting}[language=Python, numbers=none, escapeinside={(*}{*)}]
    c.copy(rank, buf, idx, (*\bfseries ch=0*)) 
    c.copy(rank, buf, idx, (*\bfseries ch=1*)) 
\end{lstlisting}

In the Hierarchical AllReduce, we manually scheduled intra-node ReduceScatters onto channel 0, the inter-node AllGathers and ReduceScatters onto channel 1, and the intra-node AllGather onto channel 2. In \autoref{fig:compiler}, this mutates the \rankdag by assigning the affected communication edges with a channel.

\paragraph{Chunk Parallelization}
\label{sec:instances}
An important aspect for performance of \SCCLang{} programs is the amount of parallelism used for transfers. Chunk parallelization is an automatic optimization that breaks up a transfer into multiple smaller transfers that execute in parallel. In the Hierarchical AllReduce, we parallelize the intra-node ReduceScatters and AllGathers using the \texttt{parallelize} modifier: 
\begin{lstlisting}
for n in range(N):
    local_ranks = [i+n*G for i in range(G)]
    with parallelize(N):
        ReduceScatter(local_ranks, 0, N)
    . . .
\end{lstlisting}
Parallelizing a code fragment by N has the effect of creating N parallel instances of 
the underlying copy and reduce operations, where each operation operates on $1/Nth$ of the data.
The compiler duplicates instruction nodes corresponding to the fragment and ensures the set channels used in each instance do not intersect so that instances execute in parallel.

There are two advantages of chunk parallelization. First, this enables parallelization of compute heavy aspects of the algorithm such as reduction. Second, parallelization can increase the utilization of high-bandwidth links by allowing multiple thread blocks to simultaneously use the underling link.
For example, our experience has shown that a single thread block in an NVIDIA A100 GPU is not capable of saturating the bandwidth of an outgoing NVLink.
The user should carefully choose the parallelization factor as increasing it beyond a certain point will reduce performance due to competition for bandwidth.

\paragraph{Aggregation.}
When the \SCCLang programs sends multiple chunks from one GPU to another and these chunks are contiguous, it is more efficient to aggregate these chunks in a single network transfer.   
Each send has a start up $\alpha$ cost and a per-byte $\beta$ cost. By aggregating sends, the compiler amortizes the start up cost. 
However, aggressive aggregation can slow down a program. All aggregated chunks must be ready before the send is executed, so that one delayed chunk blocks the progress of multiple chunks. 

Users specify aggregated sends by passing multi-count chunk references to \texttt{copy} and \texttt{reduce} operations. For example, \autoref{line:redscat_chunk} and \autoref{line:allgat_chunk} indicate that $N$ chunks are should be aggregated into a single send during the intra-node ReduceScatters and AllGathers.

\subsection{Scheduling}
Given a program's \rankdag and scheduling directives, the compiler assigns every instruction node to a thread block and remaining edges onto channels.
This assignment respects the constraints that each thread block can have at most one send and receive peer. Additionally for correctness, the assignment does not introduce deadlocks, which are possible due to the sequential execution order of instructions within a thread block.

\paragraph{Channel assignment.}
The compiler assigns every communication edge according to the user's scheduling directives. 
Any remaining edges are assigned to the lowest valid channel with two exceptions.
First, communication edges generated from a parallelized code fragment are scheduled onto replicated channels so that they don't serialize. 
Second, a series of fused instructions share the same channel. The compiler ensures this by the lowest channel that satisfies all communication edges in the chain.

\paragraph{Thread block assignment.}
\label{sec:automatic-tb}
The compiler's thread block assignment policy implements a greedy heuristic that attempts to schedule instructions in the order they will be ready. The high level steps of the routine are as follows:
\begin{enumerate}
    \item \emph{Create thread blocks:} scan through all instructions per GPU and create thread blocks for every unique (send-peer, receive-peer, channel) tuple.
    \item \emph{Calculate dependency depth:} the number of hops a chunk must traverse is used to prioritize instructions
    that are likely to be enabled earlier.\label{step:depdepth}
    \item \emph{Calculate reverse dependency depth:} instructions that contribute to a chunk that has more hops
    remaining are prioritized.\label{step:revdepdepth} 
    \item \emph{Sort instructions} into a global topological order with respect to their dependencies with a heap to prioritize nodes with a lower dependency depth first and a higher reverse dependency depth second.
    \item \emph{Assign instructions to thread blocks:} instructions are processed one by one in the topological order and
    assigned to their matching thread block. If an instruction has multiple candidates (e.g. local copies can happen on any thread block) then the one whose latest assigned instruction is earliest in the sorted order is chosen.
\end{enumerate}

All \rankdag{}s are guaranteed to have a \emph{global} topological order because it was generated by sequentially tracing the \SCCLang program. By assigning instructions to thread blocks in a topological order that respects communication and processing edges, all implicit dependencies introduced by thread block sequential execution cannot produce cycles so that the \scclef does not have deadlocks.

\paragraph{Synchronization insertion.}
An instruction's dependencies are captured in the \rankdag as communication edges between sends and receives, and processing edges that indicate execution order within a rank. The \scclef program must respect this order to be correct and data race free. While sends and receives implicitly synchronize, certain processing edges need explicit synchronization.

Instructions within a thread block are executed sequentially, and thus any processing edges between them are already satisfied. However, instruction across thread blocks execute out-of-order. Processing edges between different thread blocks are explicitly preserved in the \scclef file as cross-thread block dependencies.  

\section{\SCCLang Runtime}
\label{sec:runtime}
The \SCCLang runtime is an extension of NCCL and it inherits its infrastructure for establishing point-to-point (P2P) connections over various inter-connects including NVLink, PCIe, shared host memory, InfiniBand (IB) and TCP. \SCCLang programs are executed by an interpreter written in CUDA. All \scclef generated by our compiler is guaranteed to be correct, but some programs might only be performant for a range of buffer sizes. Therefore, the runtime dynamically selects the right algorithm to invoke based on user configurable size ranges and falls back to NCCL's built-in algorithms otherwise. This allows a user to hyper-optimize \SCCLang programs to a specific use case.

\subsection{Point-to-Point Connections}
\paragraph*{Remote Buffers}
\label{sec:remotebuffer}
NCCL abstracts different kinds of inter-connects from CUDA code by providing intermediate buffers of constant size $b$ for sends to write to and receives to read from. These buffers are subdivided into $s$ FIFO {\em slots} which allows $s$ sends to finish without waiting for receives ($1\le s\le 8$). \SCCLang compiler prevents a schedule with more than $s$ outstanding sends to avoid deadlocks. By default, $512\text{KB}\le b\le 5\text{MB}$ and $1\le s\le 8$ (exact values are defined by the protocol, explained later).

Remote buffers will be allocated on different memories depending on the inter-connection type. For NVLink or PCIe connections within a node, they are allocated on the receiving GPU. For cross-node IB connections, two buffers are allocated with one on the sender GPU and another on the receiver GPU. The IB driver transfers data between the buffers via GPUDirect RDMA~\cite{gpudirect}, with a CPU helper thread initiating RDMA transfers. Some other inter-connection types involve the host memory, but we omit their description as they are not used on our evaluation systems.

\paragraph*{Channels}
As explained in \autoref{sec:schedules}, each P2P connection in NCCL requires a \emph{channel}, which is an internal NCCL data structure that distinguishes different P2P connections between the same pair of GPUs.

\paragraph*{Protocols}
NCCL implements three communication protocols,
Simple, LL128, LL, that trade off latency and bandwidth. 
Simple has the highest bandwidth and latency, LL has the lowest bandwidth and latency, and LL128's performance is in-between~\cite{ncclrepo}. The protocol also defines the remote buffer size and the number of slots. The user
may set a desired protocol in the DSL, which is stored in the \scclef.

\subsection{Interpreter}

\begin{figure}[t]
\centering
\begin{lstlisting}[language=C++,escapechar=|]
struct Instruction { |\label{line:instr}|
  int step, opCode, srcOff, dstOff, count;
  void *srcPtr, *dstPtr;
  int depBid[D], depStep[D];
  bool hasDep; };

GC3_interpreter(Instructions instrs[N]){
  int bid=blockIdx.x; int tid=threadIdx.x;
  // chunk tiling
  for (int t=0; t < chunkSize; t += tileSize){ |\label{line:chunking}|
    // instruction loop
    for (int s=0; s<N; s++){ |\label{line:instrloop}|
      auto instr = instrs[s];
      // check for dependencies
      if (tid < D) wait(semaphore[instr.depBid[tid]], instr.depStep[tid]); |\label{line:checksema}|
      // select the instruction
      switch (instr.opCode){ |\label{line:switch}|
        case SEND:
          send(instr.srcPtr+instr.srcOff, instr.count*tileSize);
        ...}
      // set the semaphore if necessary
      if (instr.hasDep){ |\label{line:hasdep}|
        thread_fence(); sync_threads();
        if (tid==0) set(semaphore[bid],s);}}}} |\label{line:setsema}|
\end{lstlisting}
\caption{\SCCLang interpreter}
\label{fig:interpreter}
\end{figure}

\paragraph*{Initial Setup}
In the initialization phase of the runtime, all \scclef programs that will be available for use are parsed and stored in the GPU memory. If the runtime invokes the interpreter for a given program, it will launch as many thread blocks as the program needs. Note that all thread blocks in the runtime need to be running at the same time due to dependencies between them. Therefore, the compiler can only generate IRs that do not have more thread blocks than the available Streaming Multiprocessors (SMs) and the kernel is launched with \texttt{cudaLaunchCooperativeKernel} which ensures concurrent execution~\cite{cooplaunch}. The connections needed by the thread blocks in every program (see \autoref{fig:compiler}) are also initialized.

\paragraph*{Instruction Data Structure}
The execution engine for \SCCLang runtime is an efficient interpreter written in CUDA shown in \autoref{fig:interpreter} which runs a list of instructions on each thread block. \autoref{line:instr} shows the elements of an instruction: \texttt{step} is the instruction index in an array, \texttt{opcode} identifies the instruction type, \texttt{srcPtr} and \texttt{dstPtr} are the input and output pointers, and \texttt{srcOff} and \texttt{dstOff} are their corresponding offset, respectively. The pointers can be one of input, output, or scratch buffers, and offset is the chunk index into the buffer. \texttt{count} is the number of consecutive chunks this instruction will execute on (see aggregation in \autoref{sec:overview}). Last arguments are for cross thread block synchronizations: \texttt{depBid} and \texttt{depStep}. These two arrays are a list of thread block IDs and instruction steps, respectively, that this instruction is dependent on. \texttt{hasDep} is a boolean flag indicating whether there are other instruction dependent on this instruction.

\paragraph{Chunk Tiling}
\begin{figure}[t]
    \includegraphics[width=\columnwidth]{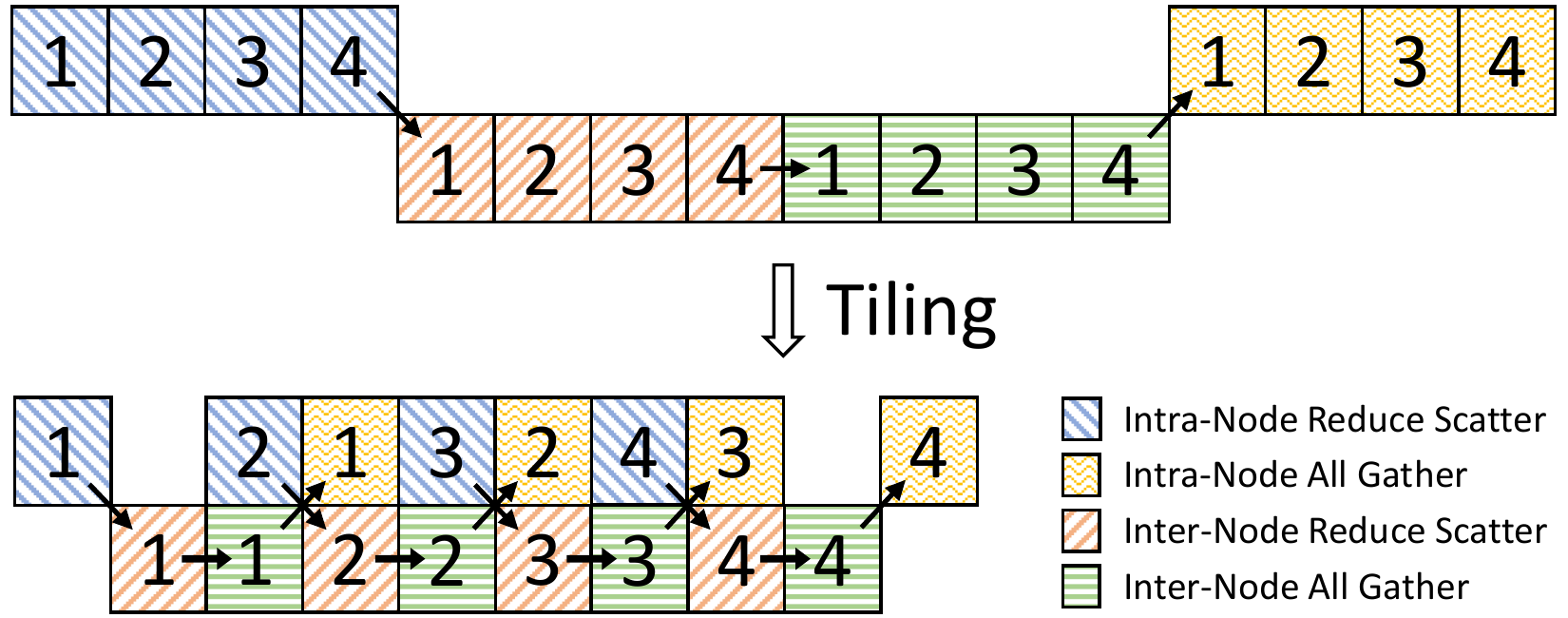}
    \caption{Pipelining induced by tiling.}
    \label{fig:hier_allreduce_tiling}
\end{figure}
The outer-most loop in the interpreter is the chunk tiling loop shown in \autoref{line:chunking} of \autoref{fig:interpreter}. As described in \autoref{sec:remotebuffer}, the remote buffers for each P2P connection have a fixed size. Therefore, if the size of each chunk is larger than the remote buffer, the runtime will automatically make small enough tiles such that each tile fits in each slot of the remote buffer. 

On the other hand, chunk tiling has a larger impact on performance as well. Consider the hierarchical AllReduce in \autoref{fig:hier_allreduce_animation}. It starts with an intra-node ReduceScatter followed by inter-node ReduceScatter and AllGather, and ends with an intra-node AllGather. The inter-node communication links are not utilized during intra-node phases and vice versa. Assuming that each phase of the algorithm is implemented on different thread blocks, chunk tiling pipelines these tiles to increase the utilization of both inter-node and intra-node communication links, as shown in \autoref{fig:hier_allreduce_tiling}. Therefore, developers may accordingly reduce the tile size for more aggressive tiling in order to maximize pipelining and performance. However, as tile sizes reduce, the additional performance benefit of pipelining decreases due to the associated startup cost of sending each message.

\paragraph*{Instruction Loop}
The inner-most loop in the interpreter in \autoref{line:instrloop} decodes instructions in the input \scclef in order. There is a list of switch-case statements in \autoref{line:switch} that decides which instructions to execute. 

\paragraph*{Cross Thread Block Synchronization}
Cross thread block synchronization is not naturally supported in CUDA. However, the interpreter runs all thread blocks
concurrently, and therefore, they can be synchronized via semaphores stored in global memory. Each thread block has a
semaphore (\texttt{semaphore[bid]}) in \autoref{fig:interpreter} which is initialized to 0 and when an instruction
\texttt{hasDep} is set (\autoref{line:hasdep}), a CUDA \texttt{\_\_syncthreads} and a \texttt{\_\_threadfence} is issued
to flush the caches and then the semaphore is set to the running step \texttt{s} (\autoref{line:setsema}). On the other
hand, if this instruction is dependent on instructions from other thread blocks, all semaphores for dependent thread
blocks wait to be set (\autoref{line:checksema}).

\section{Evaluation}
\label{sec:evaluation}
We evaluate \SCCLang by implementing classic and custom algorithms for the commonly used collectives AllReduce and AllToAll. We optimize each algorithm's schedule for various GPU system configurations and input sizes. 
All our programs require less than 30 lines of code, and took between 15 minutes to an hour to write and manually optimize.

\paragraph{Experimental Setup}
Experiments are performed on two GPU systems: type A and type B clusters\footnote{removed names for reviews}. Each node in type A contains $8$ NVIDIA A100 GPUs connected by $12$ third-generation NVLinks to $6$ NVSwitches for a total of $600$ GB/s bi-directional bandwidth. For cross-node communication, each pair of GPUs within a node share a single PCIeSwitch that connects to $2$ HDR InfiniBand NICs, each running at $25$ GB/s bandwidth. 
Each node in the type B cluster contains $16$ NVIDIA V100s divided into two boards of $8$ GPUs.  GPUs on each board are connected by $6$ second-generation NVLinks to $6$ NVSwitches, and every NVSwitch is connected by $8$ NVLinks to its counterpart NVSwitch on the other board. For cross-node communication each pair of GPUs share a single PCIeSwitch that is connected to $1$ HDR InfiniBand NIC running with $25$ GB/s bandwidth.

\SCCLang is built on top of NCCL-2.8.4-1 \cite{ncclrepo}. When applicable, we compare against collectives implemented in NCCL or expert hand-optimized implementations. 
For custom algorithms and collectives for which hand-optimized CUDA implementations were previously not available, we provide best effort hand-written kernels. 
All experiments are averaged over 50 iterations after a warmup period of 20 iterations.
Each algorithm's optimizations are tuned for the two systems. We analyze results for the A100 system since similar trends are seen on the V100 system and discuss certain exceptions. 

\begin{figure*}[t]
    \begin{subfigure}{0.45\textwidth}
        \includegraphics[width=\linewidth]{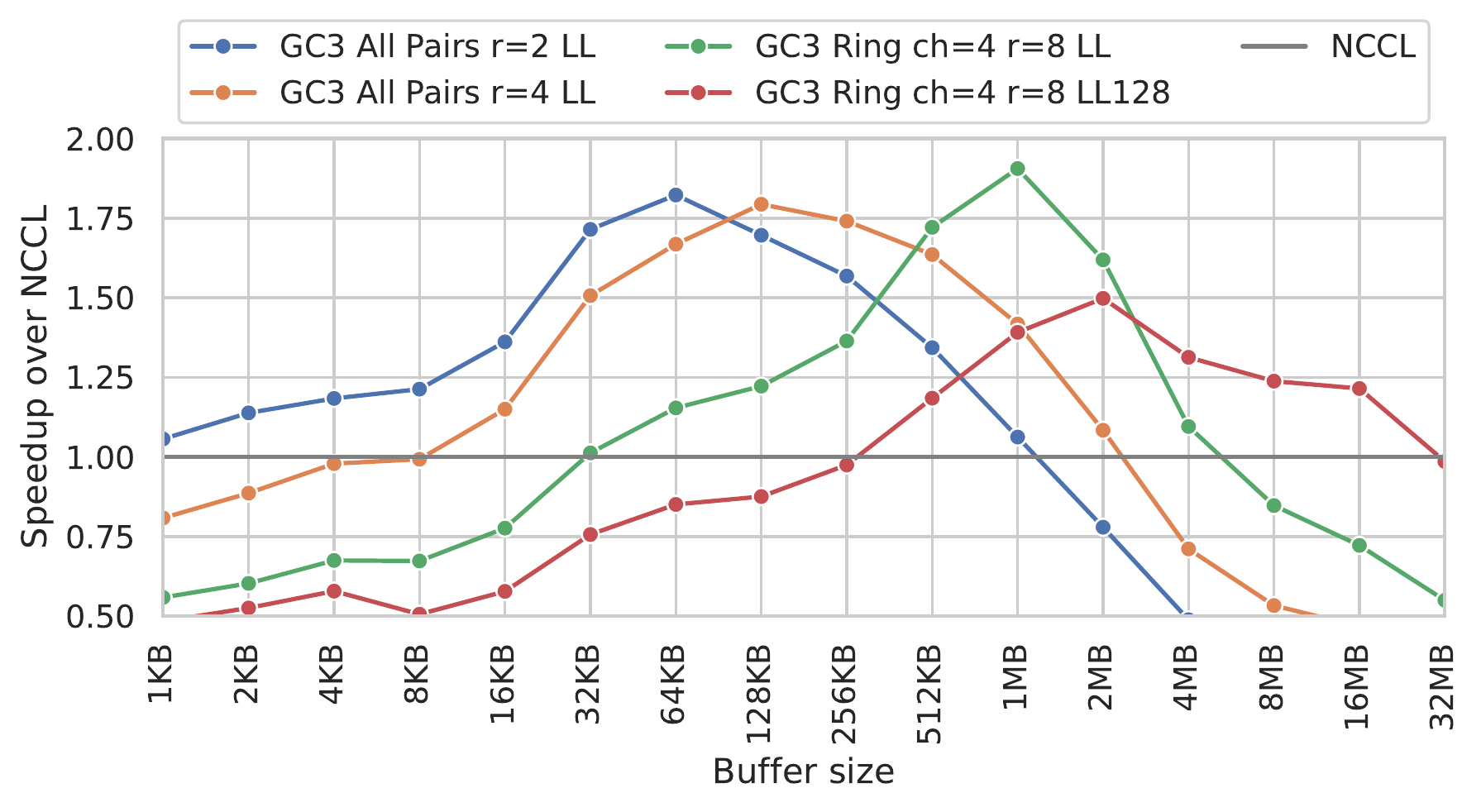}
        \caption{1 node, 8xA100 AllReduce.}
        \label{fig:allreduce_a100_perf}
    \end{subfigure}
    \hfill
    \begin{subfigure}{0.45\textwidth}
        \includegraphics[width=\linewidth]{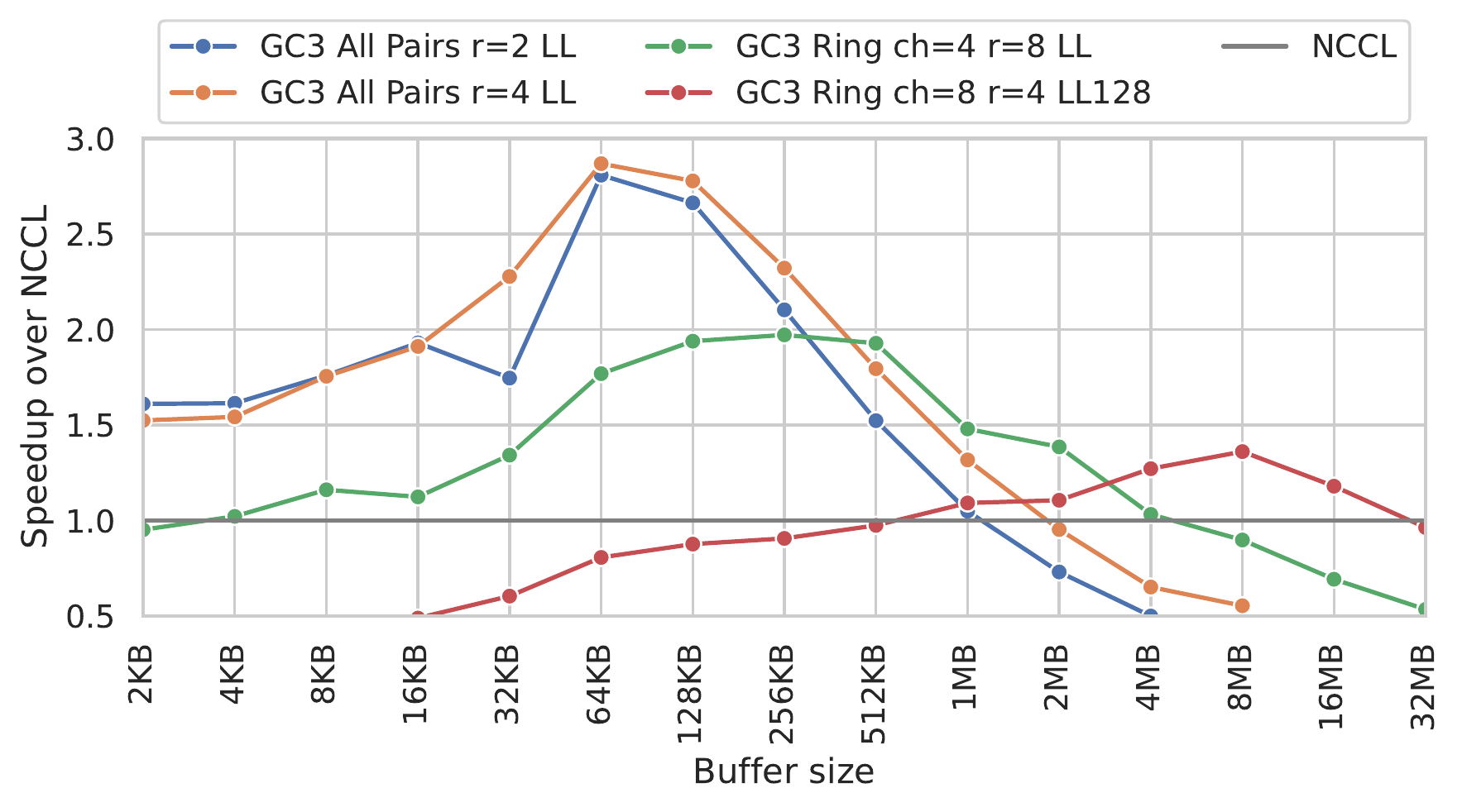}
        \caption{1 node, 16xV100 AllReduce.}
        \label{fig:allreduce_v100_perf}
    \end{subfigure} \\

    \begin{subfigure}{0.45\textwidth}
        \includegraphics[width=\linewidth]{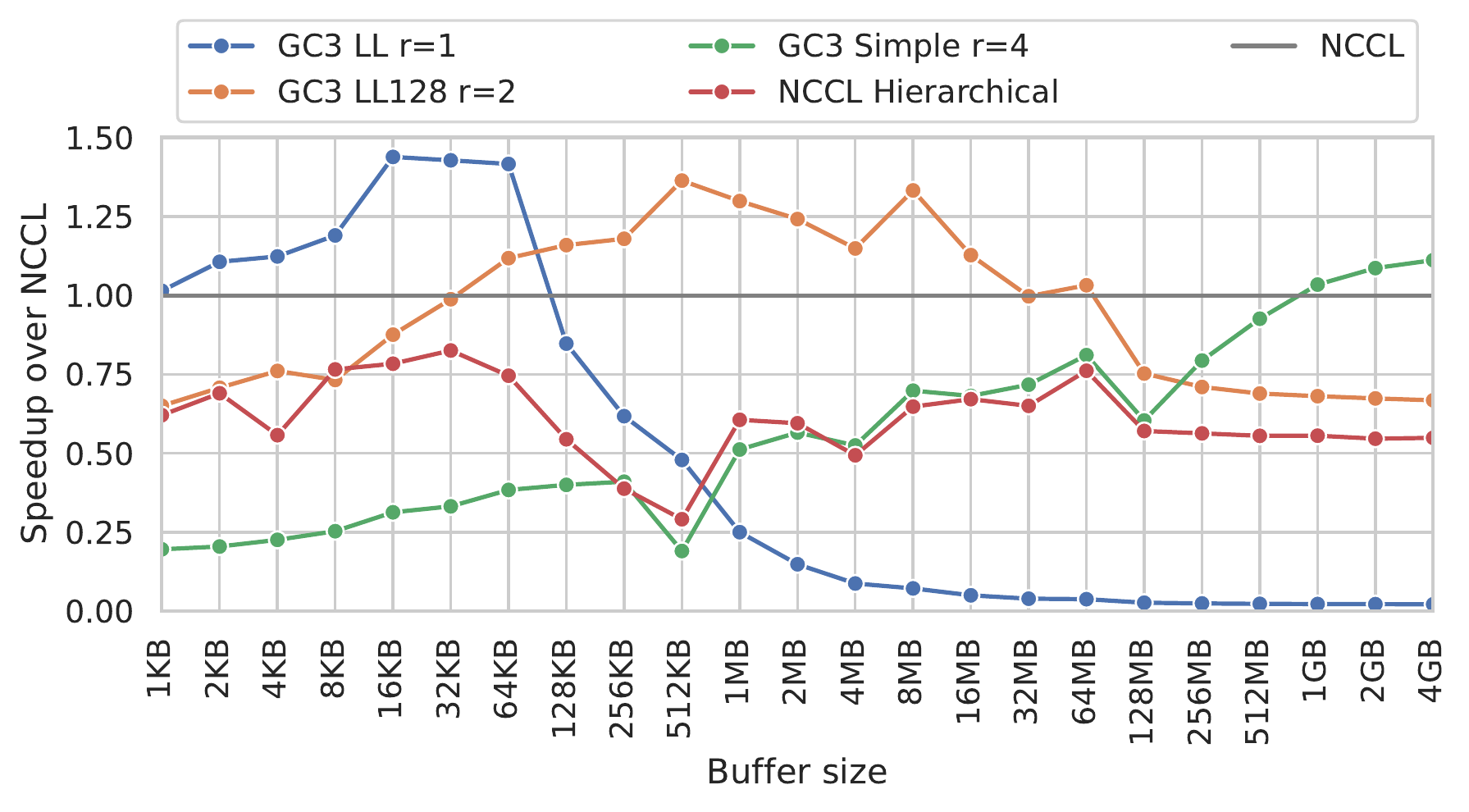}
        \caption{2 node, 16xA100 AllReduce }
        \label{fig:allreduce_hier_a100_perf}
    \end{subfigure}
    \hfill
    \begin{subfigure}{0.45\textwidth}
        \includegraphics[width=\linewidth]{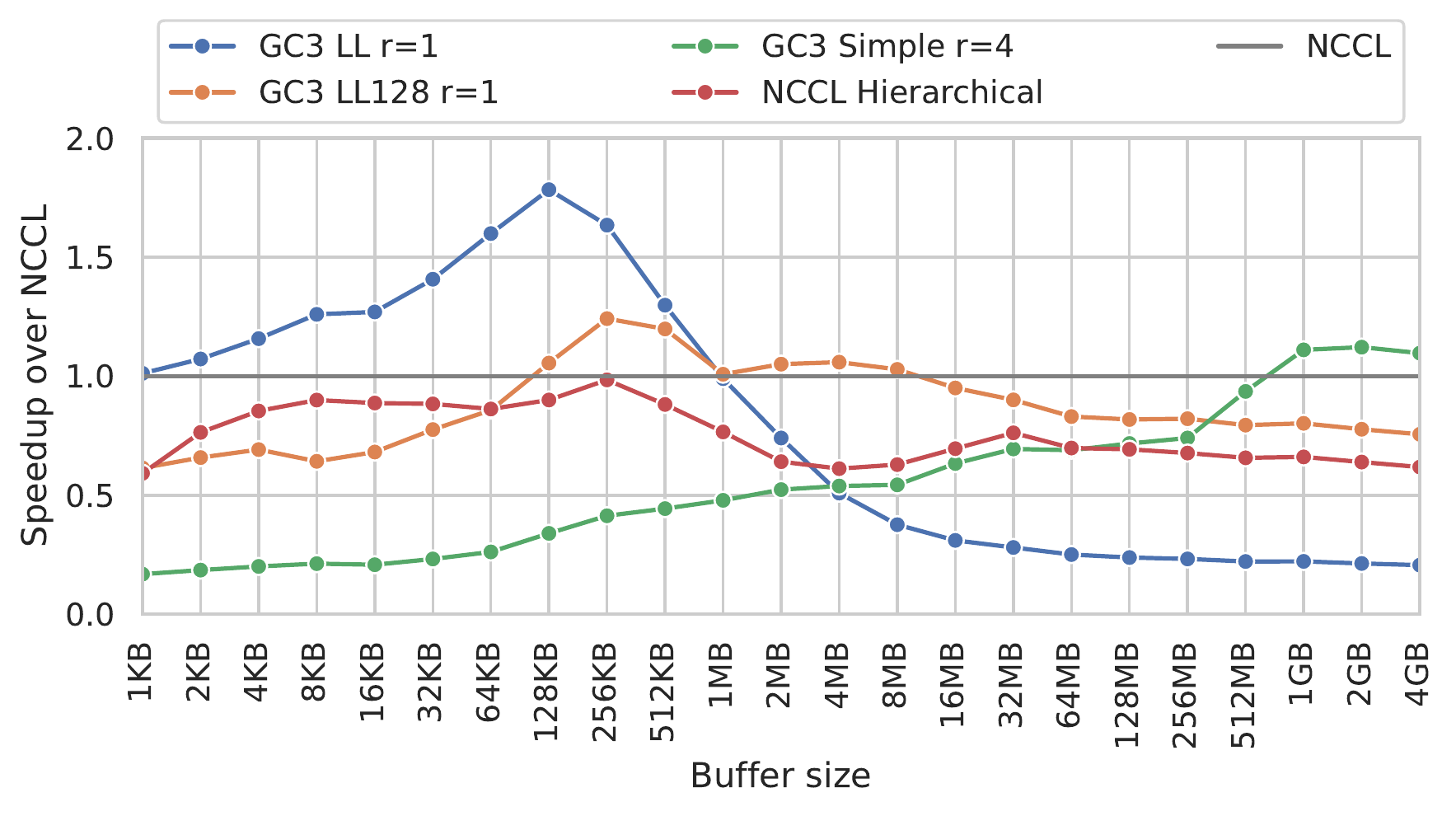}
        \caption{2 node, 32xV100 AllReduce }
        \label{fig:allreduce_hier_v100_perf}
    \end{subfigure} \\

    \begin{subfigure}{0.45\textwidth}
        \includegraphics[width=\linewidth]{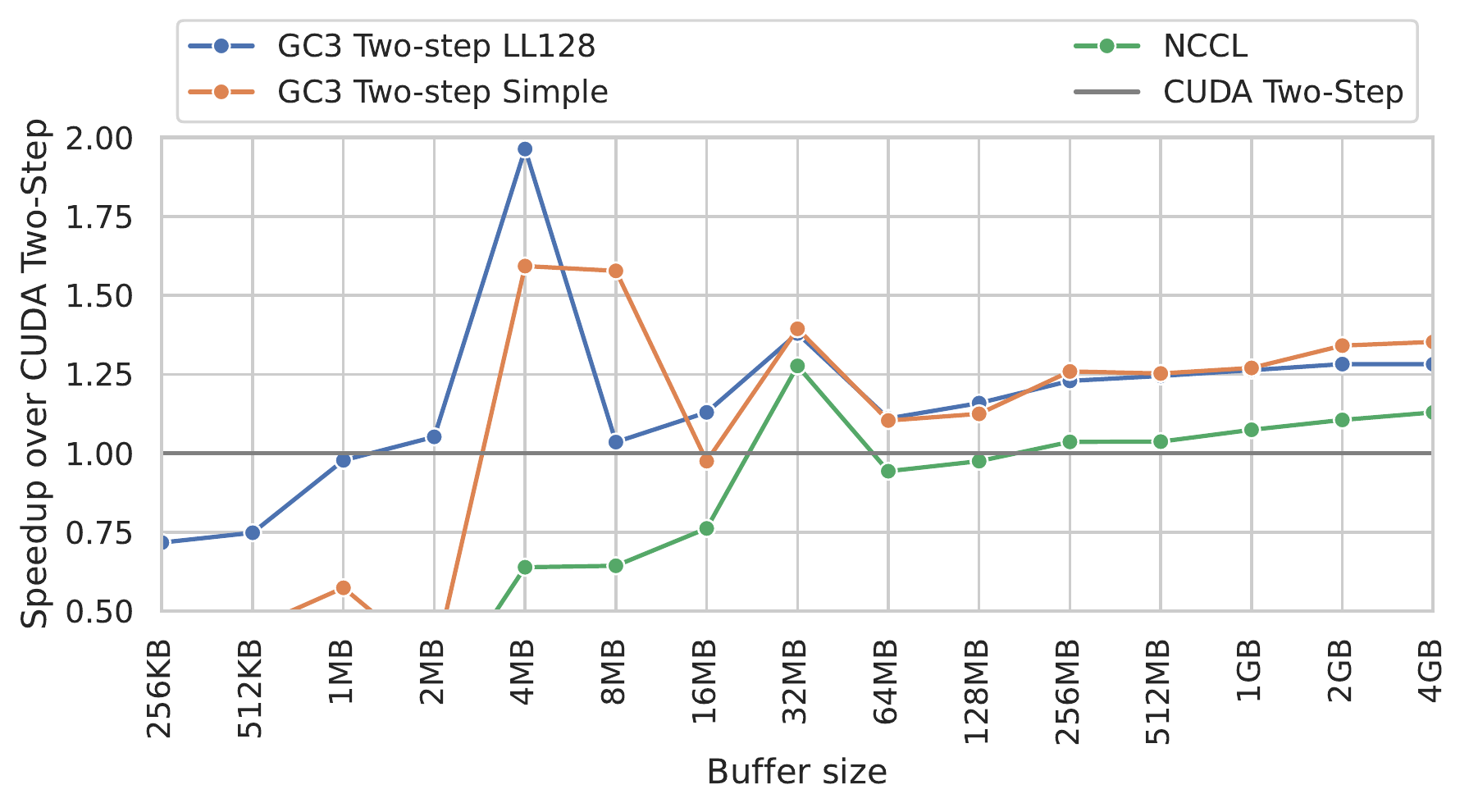}
        \caption{16 node, 256xA100s AllToAll}
        \label{fig:alltoall_multi_a100_perf}
    \end{subfigure}
    \hfill
    \begin{subfigure}{0.45\textwidth}
        \includegraphics[width=\linewidth]{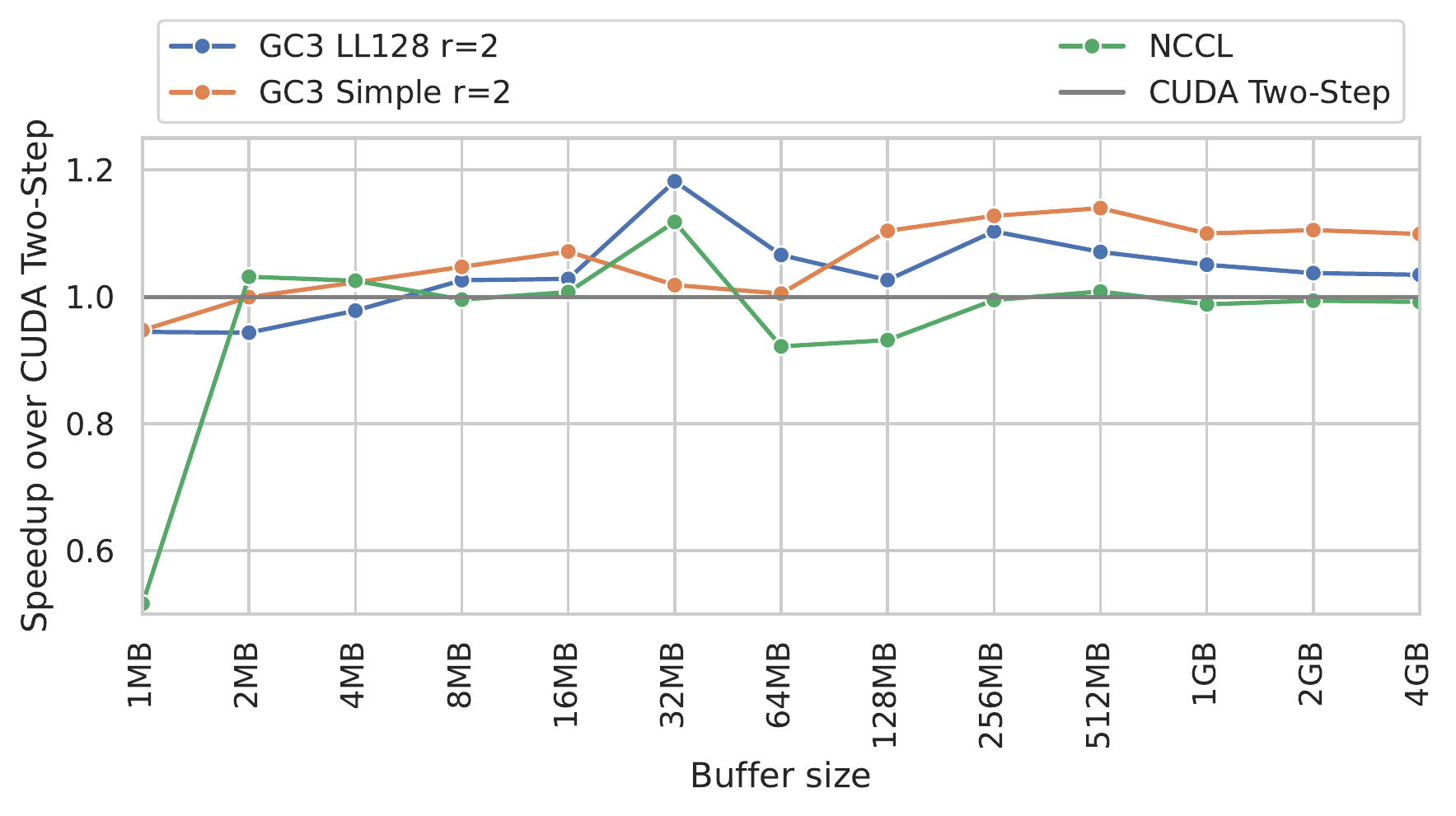}
        \caption{4 node, 64xV100 AllToAll}
        \label{fig:alltoall_multi_v100_perf}
    \end{subfigure} \\

    \begin{subfigure}{0.45\textwidth}
        \includegraphics[width=\linewidth]{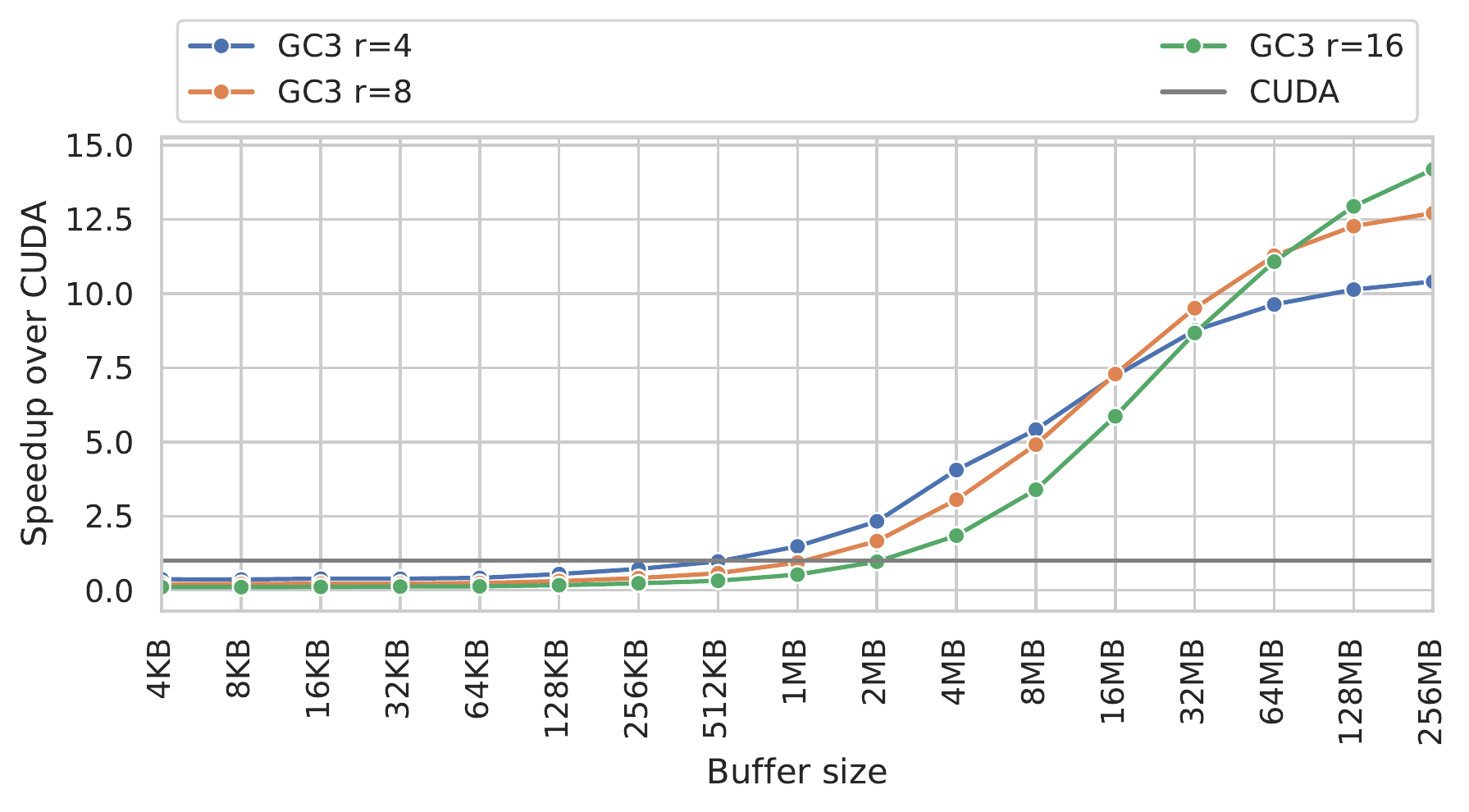}
        \caption{3 node, 24xA100 AllToNext }
        \label{fig:alltonext_a100_perf}
    \end{subfigure}
    \hfill
    \begin{subfigure}{0.45\textwidth}
        \includegraphics[width=\linewidth]{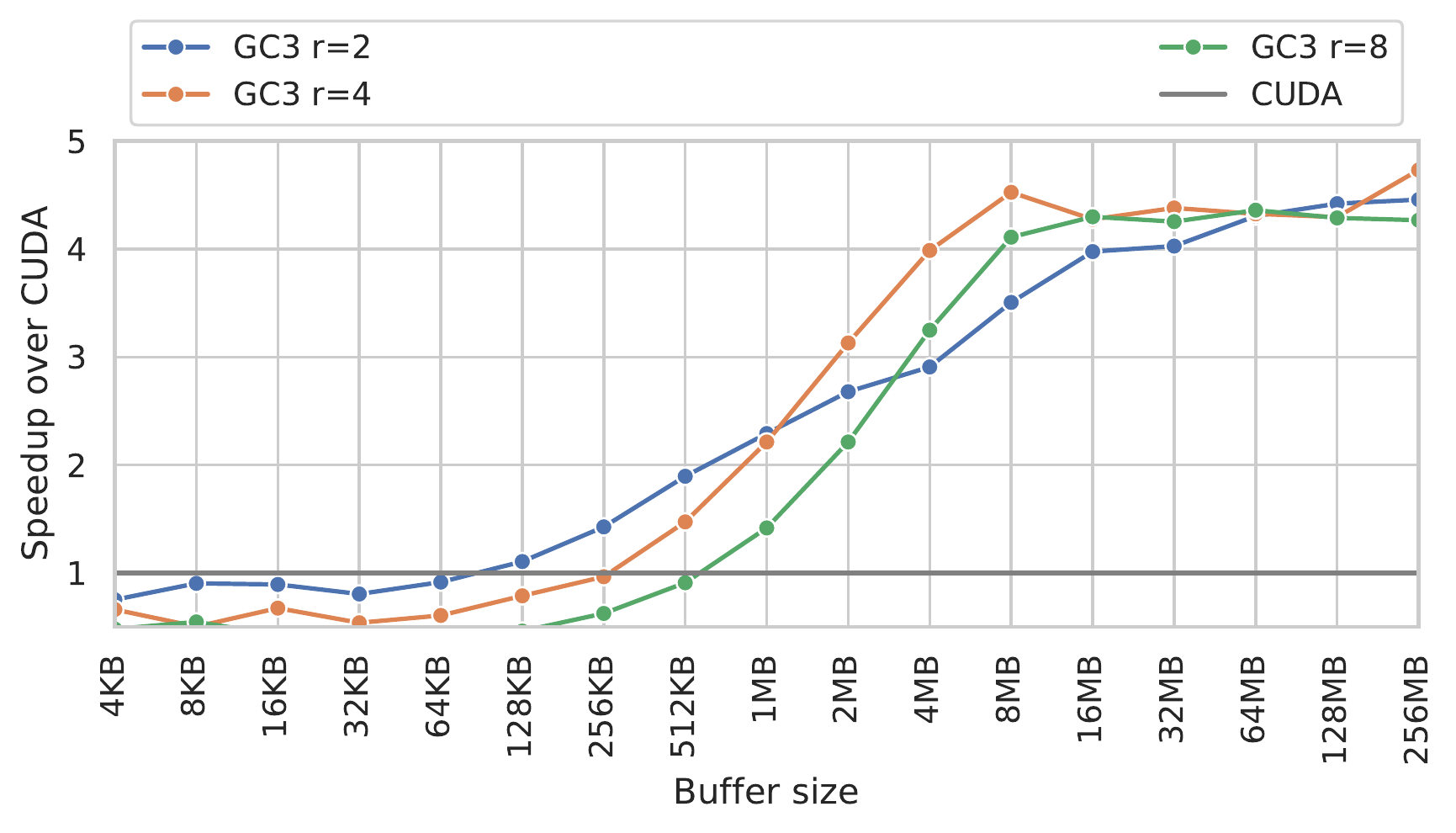}
        \caption{4 node, 64xV100 AllToNext}
        \label{fig:alltonext_v100_perf}
    \end{subfigure}
    \caption{Speedup of \SCCLang collective algorithms. Plots on the right are for A100 nodes and plots on the left for DGX2 nodes. The parameter, $r$, specifies the parallelization factor of the whole program.}
\end{figure*}

\subsection{AllReduce}
AllReduce is an MPI collective that globally reduces input buffers across GPUs and replicates the results to all GPUs. 
We implemented three AllReduce algorithms that target single-node and multi-mode systems.

\subsubsection{Ring AllReduce}

A Ring AllReduce with {\tt R} ranks, divides each rank's input buffer into {\tt R} chunks. Each chunk traverses the ring twice starting from the corresponding rank; the first traversal reduces all corresponding chunks and the second traversal copies the result to all ranks. We implement our ring with a ReduceScatter followed by an AllGather from \autoref{fig:red_scatter_all_gather} using a list of all ranks in the node, an offset of 0, and a count of 1 for the parameters.

Our ring implementation distributes a single logical ring across multiple channels by varying the channel of copy and reduce operations. We tune the number of channels per ring, parallelization, and protocol for the system.
We compare our Ring implementations against NCCL's Ring implementation in \autoref{fig:allreduce_a100_perf}.
While examining NCCL's codebase, we found and experimentally validated that NCCL's Ring schedule is roughly equivalent to scheduling a logical ring onto one channel, parallelizing the entire program 24 times, and varying the protocol based on the buffer size.

The \SCCLang ring implementation outperforms NCCL by up to $1.9\times$ when the buffer size is between 32KB and 3MB. 
Distributing a logical ring across multiple channels enables better overlapping of sends and receives resulting in performance gains. However, this distribution uses more resources thus limiting the chunk parallelization. For buffer sizes greater than 32MB, more parallelization is required, and the best \SCCLang configurations matched NCCL's performance by scheduling a logical ring onto one channel and parallelizing the program 24 times.

\subsubsection{All Pairs AllReduce}
One advantage of \SCCLang is the ability to explore different algorithms easily. 
All Pairs is an algorithm we developed while exploring algorithmic optimizations for AllReduce and is useful for small buffer sizes. This algorithm uses two communication steps: each rank receives a chunk from every rank, computes the sum, and broadcasts the chunk to every other rank. 
Since we do not have a CUDA baseline to compare against, we plot the speedup of \SCCLang's All Pairs against NCCL's Ring algorithm in \autoref{fig:allreduce_a100_perf}.

The speedups provided by All Pairs are driven by algorithmic and scheduling optimizations.
Ring and All Pairs exchange the same volume of data, but All Pairs has better latency optimality because it uses $2$ communication step compared with Ring's $2R-2$ steps. 
For buffer sizes from 1KB to 1MB, All Pairs is up to $1.8\times$ faster than NCCL, depending on the number of instances used to optimize the program. 

\subsection{Hierarchical AllReduce}

The final AllReduce algorithm we analyze is a Hierarchical AllReduce algorithm described in \autoref{sec:overview}.
\autoref{fig:allreduce_hier_a100_perf} plots the speedup of the Hierarchical AllReduce implemented in \SCCLang against NCCL. 
Depending on the input size, we apply different optimizations to the same base algorithm. 
For small sizes we are up to \allreducehierarchical faster than NCCL. For large buffers, greater than 1GB, our implementation is up to \allreducehierarchicallarge than NCCL.

In red, we plot the speedup of same algorithm implemented with NCCL collectives. The implementation is significantly slower than the \SCCLang's and NCCL's implementation due to the overhead of launching multiple kernels and lack of cross-kernel optimizations. Using \SCCLang optimizations to execute the algorithm in a single kernel and enable pipelining significantly improved the algorithm's performance such that it is faster than NCCL's implementation for a large range of buffer sizes.

\subsection{Two-Step AllToAll}
\begin{figure}[t]
    \input{text/alltoall_program.tex}
    \caption{\SCCLang Two-Step AllToAll.}
    \label{fig:alltoall_program}
\end{figure}

AllToAll is a MPI collective that transposes a buffer of data between GPUs such that chunk $i$ on GPU $j$ ends up on GPU $i$ at index $j$, and is commonly implemented as a set of point-to-point sends and receives between all GPUs.
Because each GPU exchanges data with every other GPU, AllToAll is a very communication intensive collective.

On AllToAlls spanning 10s to 100s of nodes, the naive implementation requires only one communication step, but sends many small chunks to other nodes over IB which is expensive due to the high overhead costs of IB.
We implement a Two-Step AllToAll algorithm that aggregates cross-node sends, reducing the total overhead cost of the IB sends. The algorithm is described in \autoref{fig:alltoall_program}, and we used \SCCLang's default scheduling with 1 instance and tuned the protocol for the buffer size. 

We compare the performance of our implementation against a hand-optimized CUDA implementation of the Two-Step algorithm in \autoref{fig:alltoall_multi_a100_perf}.
At large sizes the \SCCLang implementation is up to \alltoallscclvhand faster than the hand-optimized implementation. Note, at smaller sizes between 2MB-64MB there are large fluctuations in speedup caused by congestion in the IB network which is shared with other cloud tenants; however the general trends show that \SCCLang's optimizations improve performance.

For reference, we plot NCCL's performance relative to the hand-optimized implementation.
In general, both Two-Step implementation provide significant improvements over NCCL.
However, for larger buffer sizes, greater than 512MB the hand-optimized Two-Step implementation is slower than NCCL, while the \SCCLang implementation is \alltoallscclvnccl faster.

The hand-optimized version is implemented using point-to-point primitives exposed by NCCL, but lacks scheduling decisions made by the compiler that divides communication across multiple parallel thread blocks.
The \SCCLang seamless handles aggregating chunks in the scratch buffer (Line 12), while
the handwritten implementation requires a separate kernel that copies and contiguously arranges chunks in a scratch buffer for the aggregated IB send resulting in extra synchronization overhead. Furthermore, the \SCCLang implementation is much more succinct and requires only 15 lines of code while the hand optimized kernel requires roughly 70 lines of code.

\subsection{Custom Collectives: AllToNext}
A key feature of \SCCLang is the ability to define \textit{any} type of collective communication quickly and efficiently. We demonstrate this on a new collective called AllToNext. This collective involves $R$ GPUs, where GPU $i$ sends a buffer
of data to GPU $i+1$, with the last GPU sending nothing.
This communication pattern exists in applications that process data in a pipelined fashion across multiple GPUs. An obvious implementation is for every GPU to send the buffer individually. However, on a distributed system, the collective will be bottlenecked by the inter-node low-bandwidth IB links. 

We designed AllToNext algorithm specifically to distribute the inter-node transfer across multiple IB links. This is useful when there are multiple IB links per node. 
Specifically, rather than sending a single chunk through one IB link, the GPU divides the chunk and sends parts to each GPU in the node. All the GPUs in the node collectively send the parts utilizing multiple IB links.  
\autoref{fig:alltonext_a100_perf} plots the speedup of AllToNext on 3 A100 nodes against a handwritten CUDA baseline where each GPU directly sends its entire buffer to the next GPU using NCCL's send and receive primitives.

When sending small buffers, AllToNext performs worse than baseline due to overhead from the extra communication steps.
For larger buffers however, AllToNext begins to show improvement over the baseline, and is ultimately up to \alltonextspeedup for a large buffers.
The best performing selection of \textit{r} depends buffer sizes.
For small buffer sizes, less parallelization provide better performance, as the benefit from parallelizing communication doesn't offset the cost of initializing extra resources.
As the buffer sizes increase, programs with more parallelization produce larger speedups as the initialization overhead is amortized over more communication.

\subsection{End-to-End Results}
\SCCLang is currently used in inferencing a public facing language model on cloud service provider X on 8xA100 GPUs with 1.22-1.29$\times$ total GPU time speed up depending on the batch size used. \SCCLang is also used for training a large Mixture-of-Experts model for speech, language, and vision on 256xA100 GPU providing 1.10-1.89$\times$ speed up depending on the model architecture\footnote{details of these models will be revealed for the final version}.

\section{Related Work}
\label{sec:related}

The message passing interface (MPI)~\cite{dongarra2013mpi} is a popular abstraction for communication primitives. Efficient algorithms for implementing these primitives is a long-studied research area~\cite{pjevsivac2007performance, chan2007collective, thakur2005optimization}, including optimized algorithms for specific architectures like mesh, hypercube, or fat-tree~\cite{scott1991efficient,bokhari1992complete,barnett1993global} and for clusters of shared-memory processors~\cite{sistare1999optimization,traff2002improved,sanders2002hierarchical,tipparaju2003fast}.
Motivated by recent ML workloads, Horovod~\cite{alex2018horovod} implements collective primitives by using NCCL locally in node and MPI across nodes. Others such as BlueConnect~\cite{blueconnect} and PLink~\cite{plink} exploit the hierarchical network topology of a cloud system or a data center to improve the performance of collective primitives.
Recent work focuses on automatically generating new collective algorithms, either by packing trees~\cite{wang2020blink} or using a constraint solver to generate pareto-optimal algorithms~\cite{cai2021synthesizing}. In contrast, this work focuses on a high-level language for specifying these algorithms and efficiently running them on state-of-the-art accelerators. 

In-network aggregation is another direction to accelerate reduction based communication primitives.
Mellanox Scalable Hierarchical Aggregation and Reduction Protocol (SHArP)~\cite{graham2016scalable} is one of such techniques in InfiniBand switch. Other programmable switches including SwitchML~\cite{sapio2019scaling} and ATP~\cite{lao2021atp} also share the similar idea to offload GPU reduction to network switches in order to accelerate AllReduce in deep learning workloads. Apart from switches, BluesMPI~\cite{hashmi2021bluesmpi}, ACCL~\cite{he2021accl} and BytePS~\cite{jiang2020unified} also offload communication primitives to SmartNIC, FPGA, and spare CPU nodes, respectively.
Those works all introduce extra hardware thus increase bandwidth limit for primitives, while \SCCLang focuses on software stack only to program and optimize collective communication algorithms within existing hardware.

The chunk-oriented programming style of \SCCLang is motivated by {\em dataflow} programming languages. The design of the language is particularly influenced by declarative coordination languages such as Linda~\cite{Gelernter:1985:GCL} and Concurrent Collections~\cite{cnc}. Rather than use explicit tuples, \SCCLang uses implicit chunk identifiers to coordinate multiple ranks. Cilk~\cite{cilk} also influenced the aspect of \SCCLang where the deterministic semantics of the program is specified by the sequential semantics of the host language.

Recent works~\cite{zhang2017poseidon,hashemi2019tictac,jayarajan2019priority,peng2019generic} have shown the advantage of overlapping computation and communication when optimizing distributed ML workloads. While our focus here is on specifying communication collectives, extending \SCCLang to further specify the scheduling of computation is an interesting future work.

\section{Conclusion}
\label{sec:conclusion}
\SCCLang is a novel software system designed for programmable GPU collective communications.
\SCCLang provides a domain specific language for flexible collective implementations and a compiler for lowering the DSL to low-level representation which can be further executed by an optimal runtime efficiently.
We evaluated \SCCLang with common collectives AllToAll and AllReduce on different GPU systems which outperform the state-of-the-art GPU collective library, and a custom collective AllToNext which demonstrates the flexibility to program new collectives that are not in standard MPI interface.
We believe the programmability of \SCCLang will empower ML researchers to optimize existing or explore new collectives in their GPU workloads.

\bibliographystyle{plain}
\bibliography{references}

\end{document}